# The Decline of University Patenting and

# the End of the Bayh-Dole Effect

Loet Leydesdorff [a] & Martin Meyer [b]

[a] Amsterdam School of Communications Research (ASCoR), University of Amsterdam, Kloveniersburgwal 48, 1012 CX Amsterdam, The Netherlands; loet@leydesdorff.net; http://www.leydesdorff.net.

[b] SPRU, University of Sussex, Brighton, UK; Steunpunt O&O Statistieken, Katholieke Universiteit Leuven, Belgium.

**Abstract**

University patenting has been heralded as a symbol of changing relations between universities and their social environments. The Bayh-Dole Act of 1980 in the USA was eagerly promoted by the OECD as a recipe for the commercialization of university research, and the law was imitated by a number of national governments. However, since the 2000s university patenting in the most advanced economies has been on the decline both as a percentage and in absolute terms. We suggest that the institutional incentives for university patenting have disappeared with the new regime of university ranking. Patents and spin-offs are not counted in university rankings. In the new arrangements of university-industry-government relations, universities have become very responsive to changes in their relevant environments.

**Introduction**

Proponents of the Triple Helix thesis (Etzkowitz & Leydesdorff, 2000), Mode-2 (Gibbons *et al.*, 1994; Nowotny *et al.*, 2001) and the thesis of the "entrepreneurial university" (Clark, 1998; Etzkowitz, 2002) have proclaimed a shift in the function of the university and accordingly a new social contract in university-industry-government relations (Graham & Dickson, 2007; Hessels & Van Lente, 2008). University patenting has often been considered as an indicator of these developments. More recently, licensing royalties (e.g., Thursby *et al.*, 2001) and spin-off companies (e.g., Friedman & Silberman, 2003) have been added as measures of university involvement in the commercialization of technology, but the measurement of these proxies is even more complicated than patent statistics (Siegel *et al.*, 2003).

From this perspective, the Bayh-Dole Act of 1980 is often considered as a landmark in university patenting (OECD, 2000; Henderson *et al.*, 1998). This law granted permission for federally funded researchers to file for patents, and to issue licenses for these patents to other parties. However, Mowery & Sampat (2005) have argued that the law can be considered as both an effect and a cause of increased university patenting before and after its passage, and they have charted (Figure 1) the continuously increasing participation of US universities in the national patenting system since 1963.

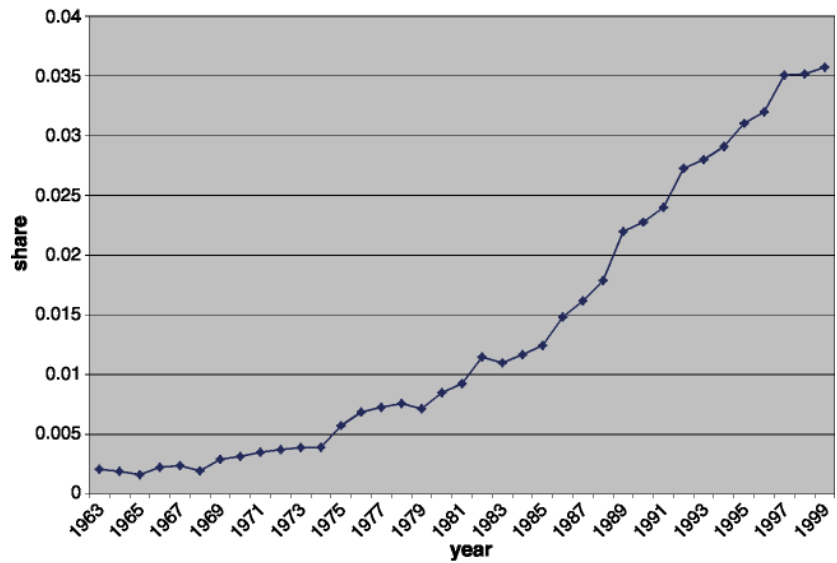

**Figure 1**: US research university patents as a percentage of all domestic-assignee US patents, 1963-1999. Source: Mowery & Sampat, 2005, at p. 120).

The proclaimed effects of the Bayh-Dole Act on university patenting in the USA have encouraged other governments to introduce similar legislation. For example, in Belgium the corresponding Flemish regulations emphasize the importance of research in relation to commercialization. Recent Finnish legislation distinguishes between inventions resulting from open innovation and collaborative efforts between universities and third parties (Meyer, 2008). However, IP regulations vary among countries and regions. Among the new member states in Central Europe, Slovenia and Hungary, for example, have adopted Bayh-Dole-style laws while others have not.

Mendes & Liyanage (2002) reported that Australian universities emulate what they perceive as the Bayh-Dole success even without legislation. However, these efforts have sometimes had only marginal success. In a recent review of the contributions of Italian universities to the processes of technology transfer and commercialization, Baldini *et al.* (2006 and 2007) came to a similar conclusion: university patenting and related activities need a fertile context to develop both inside and outside the campus. The US success story cannot be imitated simply by changing IP laws and by transferring ownership. Von Falck & Schmaltz (2005) reported that the change in the German legislation led to problems in developing new collaborations between universities and industries. Using statistics, however, Van Looy *et al*. (2007) showed that the introduction of Bayh-Dole type legislation had an independent effect on patenting by universities when compared among European nations. The increases ranged from 250% for Germany, or 300% for Belgium, to 500% for Denmark.

We have noted for some time that the Bayh-Dole effect in the USA itself has withered away, with a relative decline of university patenting since 2000. However, since our indicators were not sufficiently robust, we have not previously published these results. More recently, Wong &Singh (2007) published data about the numbers of US patents by leading universities as a percentage of all patents in the database of the US Patent and Trade Office (USPTO). This data, and the data made available by the Association of University Technology Managers (AUTM, 2008) in their yearly Surveys of US Licensing Activity, corresponded so well with our previous results that we investigated the noted decline of university patenting further.

**Methods and materials**

Three international databases of patents are fully searchable online in English: the American USPTO at http://www.uspto.gov, the database of the World Intellectual Property Organization (WIPO) at http://www.wipo.org, and the search portal of the European Patent Organization (EPO) at Esp@cenet at http://ep.espacenet.com/advancedSearch?locale=en_EP. This last database offers three search options: the European database, the WIPO database, or worldwide searching.

Criscuolo (2006) discussed a so-called 'home advantage effect' of patenting. This means that one can expect patents to be overrepresented in their country of origin. This may be gradually changing in the European Union, where inventors now have several options for filing patents: at the national level, the European office, or the WIPO (Leydesdorff, 2008). The various routes have different advantages and disadvantages (Dolfsma & Leydesdorff, 2008). Perhaps one can expect universities first to patent at home more than industries do. In any case, the search portal of the EPO at Esp@cenet is most convenient to use since one can search for the patent portfolios of specific universities worldwide.

The USPTO database was searched for each year with the word "university" in the name of the applicant. The results underestimate the number of applications by universities because some universities may have names like the Massachusetts Institute of Technology (MIT) which cannot be found with this strategy. Statistically, however, searching with "Institute of Technology"— including all institutes with this name (e.g., the Indian Institute of Technology)—generate a recall an order of magnitude lower than the search for the term "university".

The stability of the procedures in the USPTO as a national domain provides us with an advantage above international databases like the WIPO, which include very dynamic environments such as China and India. In these rapidly changing environments, several dynamics may interact. The EPO database, for example, jumped from 19,876 patents published in 2005 to 126,611 in 2006, indicating that this database is still under construction. While the WIPO database is stable, it suffers from a rapid expansion because of its most international character.

The searches are based on publication years of issued patents because patent applications have been published by the USPTO only since 2001. According to the statistics provided by the AUTM (2008, at p. 28), US universities would file patents more than three times as often as they are granted. Inclusion in the database may lag behind the publication year of the patents, but these effects are mainly important for the most recent year. All searches for the period 2000-2007 were therefore repeated in January, 2009. The numbers for 2008 may be underestimated in the case of the EPO and WIPO databases.

**Results**

Our main results are summarized in Figure 1. This figure is based on three independent sources: the squares (■) indicate the number of university patents (normalized as a percentage of USPTO patents) as listed in the yearly reports of the Association of University Technology Managers (AUTM, 2008); the diamonds (♦) indicate this percentage as measured by searching with the word "university" in the field of the assignees among the patents issued during the period 1977-2007. Wong & Singh (2007) provided numbers for university patenting in the USPTO database (● ; all universities).

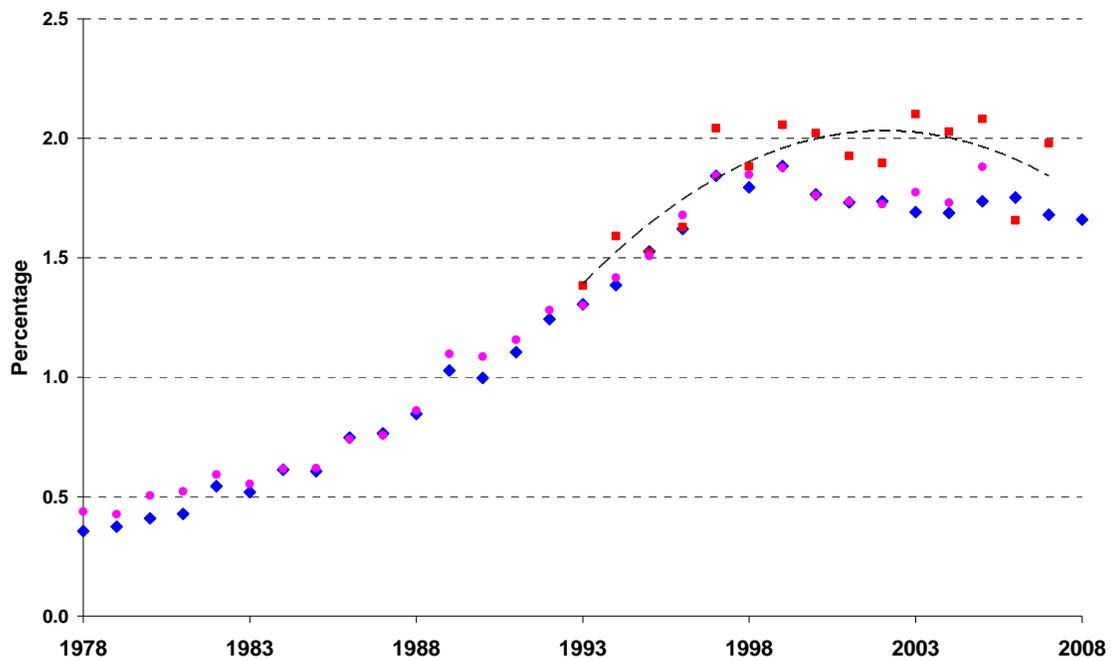

**Figure 1**: University patenting (1978-2008) as a percentage of patenting in the USPTO database. (Sources: ■ AUTM, 2008; ♦ online search at <u>http://www.uspto.gov</u>, 15 January 2009; ● Wong & Singh (2007).)

The three lines match in terms of the trends. As noted, our searches with "university" as word in the names of the applicants underestimate the total numbers and the line is therefore the lowest one. Wong & Singh (2007), however, excluded the University of California from their data because it is not possible to distinguish between the eleven campuses of this university in terms of the patent registrations. The AUTM data is probably most comprehensive, but the time series is limited. Given the correspondence among the lines, however, we submit that the stabilizing and somewhat downward trend is robust. Note that we use issued patents, since the number of applications per year by US universities can be more than three times higher (AUTM, 2008, at p. 28).

As noted, our online indicator omits institutes like MIT that do not have the word "university" in their name. For this reason, Figure 2 provides the results of searching the Esp@ce database for worldwide patenting in four major American universities, among which MIT and CalTech.

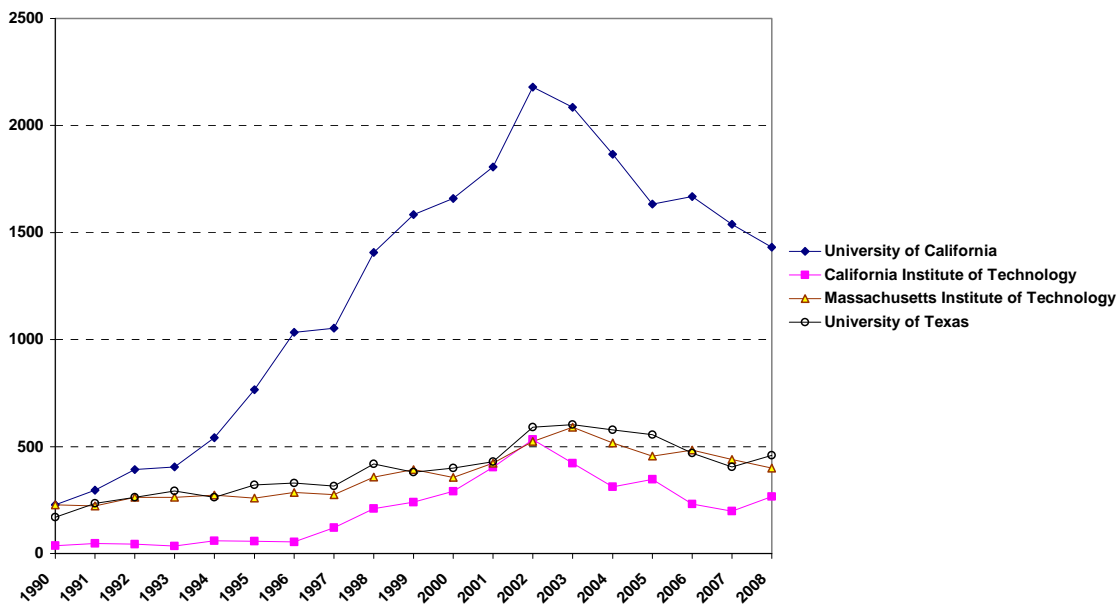

**Figure 2**: Worldwide patents of four leading US universities. (Source: Esp@ce database; 15 January 2009)

Figure 2 first shows that worldwide patenting in US universities is now at a considerably higher level than domestic patenting despite the "home advantage" effect (Criscuolo, 2006). Within the USPTO database, the University of California—which is an aggregate of eleven universities in California—peaked with 468 patents in 2002, while 2,230 patents could be counted worldwide as the peak in this same year.

Whatever the measurement problems with these different databases may be, the trend is clear and not exclusively American. Figure 3 provides a figure in the same format for four leading non-American universities. (ISIS Innovation was added to the graph for Oxford University because the university uses the services of this bureau for its patenting.)

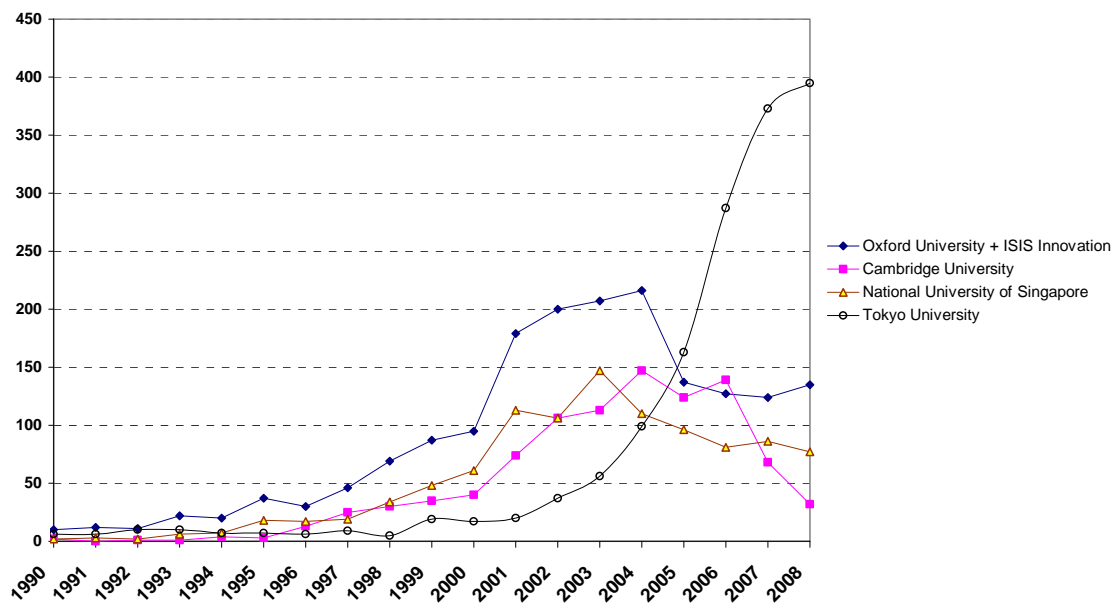

**Figure 3**: Patenting by leading non-American universities.[1] (Source: Esp@ce database; 7 June 2008).

While all curves exhibit stabilization or decline, Tokyo University is the single exception with an ongoing increase in patenting at an exponential rate. This is probably caused by strong incentives from the national government. Note that patenting by European universities took off only during the 1990s, while the American universities had already increased their (mainly domestic) patenting activities during the 1980s.

---

[1] Numbers for Tokyo University are based on adding "Tokyo University" and "University of Tokyo" as search terms.

**Discussion and Conclusions**

At the global level university patenting is still gaining momentum, but in the most advanced economies the effects of the Bayh-Dole Act of 1980 seem to have faded away since the turn of the millennium. In our opinion, the reason for this is structural. More universities are nowadays increasingly ranked in terms of their knowledge output, and patents or spin-offs are usually not part of this ranking (e.g., THES, 2008). The nature of the competition among universities is changing, and the incentive to patent has thus withered. International collaborations and coauthorships, for example, have become more important in research assessment exercises than university-industry relations (Glänzel, 2001; Leydesdorff & Sun, 2009; Persson *et al*., 2004; Wagner, 2008).

When we presented these results at conferences—with technology transfer officers and researchers who patented among our audience—the main counter-argument was that the observed decline would be the effect of "institutional learning" by universities. However, why would it have taken American universities twenty years to learn that university patenting is expensive and not always rewarding, while this problem was noted extensively in the relevant literature during the 1990s (Rosenberg & Nelson, 1994; Webster & Packer, 1997; Rappert *et al*., 1999)? Why would rising costs of patenting be prohibitive, while the number of staff members in technology transfer offices continues to increase with more than 5% per year (AUTM, 2008, at p. 19). Patenting has always been expensive, particularly when pursued internationally. As we showed above, university patenting is declining both domestically and internationally.

More recently, the number of spin-off companies from academic institutions has also declined (Mustar, 2007). Furthermore, this author noted that university incubators entertain decreasing links with the research process itself. Other opponents noted that universities may increasingly be inclined to outsource patenting. However, we included Oxford University which outsources most of its patents through ISIS Innovation. In this case, one can also see a sharp decline since 2004.

In our opinion, these developments can also be appreciated differently. The return of universities to core missions does not imply that the Triple Helix thesis has lost its validity: the system has changed by engaging with its relevant environments no longer in terms of institutional boundaries, but increasingly in terms of functional relations. These are manifested in neo-institutional arrangements that can be shaped and dissolved in collaborations and competitions much more flexibly than before. Thus, a new social contract has been shaped between academia and industries and governments as the main partners in the production of knowledge. Patenting, however, has become a possible function of universities, albeit not a core one, as the proponents of Mode-2 and the institutional version of the Triple Helix thesis predicted. The third mission has remained a latent one, including new forms of education, incubation, and long-term commitments to social values.